\documentclass[a4paper]{article}

\usepackage{float}
\usepackage[
  a4paper,top=3cm,bottom=2cm,left=3cm,right=3cm,marginparwidth=1.75cm]{geometry}
\usepackage{amsmath}
 
\usepackage{amsthm,bm,mathtools,etoolbox,amsfonts,bbm}
\usepackage{hyperref}
\hypersetup{colorlinks,linkcolor={blue},citecolor={blue},urlcolor={blue}}    
\usepackage[nameinlink,capitalise]{cleveref}
\usepackage[utf8]{inputenc} 
\usepackage[T1]{fontenc}    
\usepackage{url} 
\usepackage{booktabs}
\usepackage{siunitx}
\usepackage{caption}
\usepackage{natbib}
\usepackage{placeins}
\usepackage{adjustbox}
\usepackage{setspace}
\usepackage{algorithm}
\usepackage{algpseudocode}
\usepackage{multicol} 
\usepackage[dvipsnames]{xcolor}
\usepackage{makecell, multirow, tabularx} 
\usepackage{rotating}
\usepackage{float}
\usepackage{ulem}
\usepackage{graphicx}
\usepackage{tabularx}
\usepackage{authblk}
\usepackage{svg}
\usepackage{soul}
\usepackage[title]{appendix}

\newtheorem{Remark}{Remark}[section]

\newcommand{\notcheckmark}{{$\surd$}\textsuperscript{\textcolor{black}{\kern-0.35em{\bf--}}}}

\usepackage{tikz}
\usepackage{circuitikz}
\usetikzlibrary{shapes.geometric, arrows}

\title{Prediction Inference for Hybrid Models on Realized Volatility Forecasting}

\author[1]{Kyle Parrella}
\author[1]{Kejin Wu}

\affil[1]{Loyola University Chicago, Department of Mathematics and Statistics}

\date{}

\begin{document}

\maketitle

\begin{abstract}

\noindent GARCH-LSTM hybrid models have become increasingly popular in the volatility forecasting literature. To date, these models have almost exclusively focused on point estimation and accuracy measures such as RMSE and MAPE, while failing to assess the uncertainty of these point estimates. We approach this gap in two ways. First, we extend the existing GARCH-LSTM model by modeling the conditional mean with an ARMA structure. Second, we develop a forward bootstrap algorithm to build multi-step-ahead prediction intervals for future realized volatility. In order to address the notorious undercoverage of prediction intervals, we implement two solutions. One draws bootstrap innovations from the empirical residual distribution to mimic the fat-tail property of financial series. The other utilizes principles from conformal prediction to systematically extend the prediction interval length. We evaluate the methods in simulations with heavy-tailed and time-varying GARCH processes and in rolling-window forecasts of four U.S. stock indices from 2020 to 2025. The ARMA-based hybrid gives the lowest point-forecast errors in most settings, and 
combining both remedies with the mean-based hybrid yields intervals closest to the nominal level among competing methods.

 
\end{abstract}

\section{Introduction}\label{Sec:Intro}

Volatility forecasting plays a critical role in many fields of finance, such as derivative pricing and portfolio risk management. As a result, reliable and accurate methods for volatility forecasting are essential to the success of these industries. Traditionally, these markets have used the econometric ARCH and GARCH-type models proposed by \cite{engle1982arch} and \cite{bollerslev1986garch} to meet this need. Over the years, variations of these models, such as the EGARCH \citep{Nelson1991EGARCH} and GJR-GARCH models \citep{glosten1993}, have evolved from their predecessors as improved methods in certain use cases. Similarly, the EWMA model was originated with \cite{Roberts1959} and further refined by \cite{longerstaey1996riskmetricstm}; see \cite{francq2019garch} for a comprehensive introduction of GARCH-type models. 

In response to the limitations of traditional econometric techniques for volatility forecasting, researchers have turned to machine learning techniques because of their ability to better capture nonlinear and nonparametric trends. It is well-known that neural networks have the universal approximation property, i.e., they can approximate any functions under some smoothness conditions; see \cite{mcculloch1943logical} and \cite{cybenko1989approximation,hornik1989multilayer,yarotsky2018optimal, yarotsky2020phase} for this property displayed in simple,  multilayer and deep feed-forward neural networks, respectively. Motivated by this powerful estimation ability, researchers have increasingly applied these neural network-type models to a wide range of scientific areas, including finance. In financial mathematics, researchers derive the prediction of volatility directly from a well-trained neural network and prediction is conducted a priori. One unfortunate feature of the neural networks in this application is that they function as a sort of ``black box'', which makes statistical reasoning between predictions and historical series difficult. Instead of purely relying on neural networks, \cite{hajizadeh2012hybrid} took the output from standard GARCH models as the input of neural networks to improve the accuracy of forecasting volatility, resulting in a so-called hybrid model. Similar works include those of \cite{garciamedina2023} and \cite{wanglu2024}. \cite{kim2018forecasting} proposed the idea of feeding the estimated parameters of different types of GARCH models with other explanatory variables as input to the Long Short-Term Memory (LSTM) model, a form of recurrent neural networks originally proposed by \cite{schmidhuber1997lstm}. LSTM models have modeling applications in a variety of fields including physics, chemistry, and finance; see \cite{Zhang2020}, \cite{Mandis2024} and \cite{Moghar2020} for more details, respectively. 

GARCH-LSTM models have become increasingly common in volatility forecasting literature. This is due to the theoretical and empirical advantages these models have \citep{kim2018forecasting,garciamedina2023}. The fundamental idea of the hybrid proposed by \cite{kim2018forecasting} is that the LSTM is exceptional in leveraging the high-level temporal patterns of the time series, which are revealed by the parameters of various GARCH models. By combining the networks and classical models, hybrids are able to capture the econometric and deep learning techniques each model presents individually. GARCH creates a structured framework for defining conditional volatility, while LSTM is able to capture long-range and nonlinear interactions that can be missed by GARCH. Hybrid models also offer computational benefits, as they can be less computationally intensive than strict LSTM modeling \citep{wanglu2024}.

However, the existing literature on this topic emphasizes point estimation and improving empirical accuracy metrics such as root mean squared error (RMSE) and mean absolute percentage error (MAPE). There is a lack of literature that references more rigorous statistical measures, such as prediction intervals or uncertainty measures. 
Motivated by the forward bootstrap prediction methods developed gradually in \cite{pan2016bootstrap,politis2023multi,wu2024bootstrap}, we propose a form of forward bootstrap algorithm to build prediction intervals for volatility forecasting involving LSTM. In short, the forward bootstrap prediction technique offers an approach to evaluate the whole distribution of forecasting. We additionally propose a systematic way inspired by the conformal prediction idea of \cite{vovk2005algorithmic} to widen the prediction intervals, improving the coverage rate for finite sample cases.  Similar to the usage of a confidence interval in quantifying the accuracy of parameter estimations, the prediction interval can be utilized as a complementary tool to guide the decisions of economists on determining the prices of derivative goods, the risk of assets and the allocation of portfolios, etc. Moreover, we further improve the current GARCH-LSTM model in \cite{kim2018forecasting} by modeling the mean component of the financial series with an ARMA model; consistent with the idea of ARMA-GARCH modeling. Our contributions are summarized as follows in Section \ref{Subsec:contributions}.

\subsection{Our contributions}\label{Subsec:contributions}
\begin{itemize}
    \item We refine the hybrid model in the seminal work \cite{kim2018forecasting} to forecast realized volatility. More specifically, we incorporate a machine learning model with the ARMA-GARCH framework to perform volatility forecasting. To train the model, we define a specific MSE loss criterion that trains an LSTM model to approximate the conditional variance when the true variance is not observed. Moreover, we propose a maximum likelihood estimation algorithm to train all parameters of the LSTM and the ARMA-GARCH framework simultaneously.
    \item Beyond point prediction, we propose the forward bootstrap algorithm to make (multiple-step-ahead) prediction intervals for realized volatility. Additionally, we further provide two strategies to improve the coverage rate of the prediction interval in finite sample cases: (1) we propose
    a systematic way to extend the prediction interval using the principles of conformal prediction; (2) we use empirical residuals, instead of the normally distributed error usually assumed in the GARCH model, to ignite the forward bootstrap procedure for the construction of prediction intervals with a hybrid model. This modification emulates the fat-tail property of financial returns. It is also of note that our forward bootstrap algorithm can be applied to naive LSTM or other model-based machine-learning methods.
    \item Comprehensive simulation and real data analyses are deployed to compare our methods with classical models. Several empirical findings are clear: (1) the hybrid LSTM models with the additional information from the estimated variance and coefficients of classical models have a better point prediction performance compared to naive LSTM and classical models; (2) the naive LSTM method can be beaten even by classical models for some simulations and real data scenarios; (3) even predictions over long horizons perform remarkably well, with the hybrid models outperforming classical models in coverage on horizons of 20 days (see Tables~\ref{Table:M1H20} and~\ref{Table:M2H20}).
    
\end{itemize}

The paper is organized as follows. The structure of the hybrid model is explained in Section \ref{Sec:HModel}. Then, our algorithm to construct the prediction interval is illustrated in Section \ref{Sec: BootPI}. Simulations and real-data analysis are given in Sections \ref{Sec:simulation} and \ref{Sec:realdata}, respectively. Section \ref{conclusions} concludes this work.

\section{Hybrid Model}\label{Sec:HModel}
\subsection{GARCH model}\label{Subsec:GARCH}

The ARCH model, introduced by \cite{engle1982arch}, allowed for the modeling of conditional variances, which are generally common for real-world financial time series data. Subsequently, the GARCH model was introduced by \cite{bollerslev1986garch} to improve the generalizability and provide a more flexible lag structure of the ARCH model.  

The GARCH(p,q) process is used to model the future variance by considering the past $q$ variances and $p$ error terms. We define $\varepsilon_t$ to be a real-valued discrete-time stochastic process and $\mathcal{F}_t$ to be the information set of all information available through time $t$. The GARCH($p$,$q$) process is then given by $$\varepsilon_t|\mathcal{F}_{t-1} \sim N\left(0, \sigma^2_t\right),$$
$$\sigma^2_t= \alpha_{0} +\sum_{i=1}^p \alpha_{i}\varepsilon_{t-i}^2+\sum_{j=1}^q \beta_{j} \sigma^2_{t-j}.$$

\subsection{EGARCH model}\label{Subsec:EGARCH}
While innovative and more practical than ARCH models, the GARCH($p$,$q$) process was flawed in certain applications. \cite{Nelson1991EGARCH} identified and reconciled three major flaws. The first of these issues was that the GARCH($p$,$q$) process failed to capture some trends that were observed in financial data. Namely, the negative correlation of stock returns and changes in returns volatility that was observed by researchers was not captured in the model. Secondly, it was difficult to interpret the ``persistence'' of shocks to the conditional variance. The final flaw identified was that the non-negativity constraints of certain variables were restrictive in practice. 

To reconcile these flaws, \cite{Nelson1991EGARCH} proposed a new GARCH mechanism known as exponential GARCH (EGARCH). It is obvious that if $\sigma_t^2$ is the conditional variance given $\mathcal{F}_{t-1}$, i.e., the information at time $t-1$, then it must be nonnegative with probability one. As opposed to other GARCH models that ensure nonnegativity by making $\sigma^2_t$ a linear combination of positive random variables, \cite{Nelson1991EGARCH} adopted a different function to ensure $\sigma^2_t$ remained nonnegative: $$ \ln(\sigma_t^2) = \alpha_0
+ \sum_{i=1}^{p}\Big[\alpha_i\big(|z_{t-i}| - \mathbb{E}|z_{t-i}|\big) + \theta_i\, z_{t-i}\Big]
+ \sum_{j=1}^{q}\gamma_j \ln(\sigma_{t-j}^2), $$
where $$z_t= \varepsilon_t/\sqrt{\sigma^2_t}.$$

\subsection{EWMA model}\label{Subsec:EWMA}The Exponentially Weighted Moving Average (EWMA) is a statistic that represents the average of a value of interest over a given time period of length $T$ where the values of earlier inputs are given less weight in the EWMA value. By including a weighting factor $\lambda$, the EWMA's sensitivity to drift can be adjusted. Our definition of the EWMA process originates from \cite{Roberts1959} and is as follows: 
$$\mathcal{S}_t = \lambda \mathcal{S}_{t-1} + (1-\lambda)Y_t,$$ 
where $t=1,2,..., T$; $Y_t$ is the point observation at time $t$, and $\mathcal{S}_t$ is the weighted average of interest a time $t$. By the definition above, we can determine the EWMA process's ``depth,'' that is, how much past observations impact $\mathcal{S}_t$,  by adjusting $\lambda \in(0,1]$, where the larger $\lambda$ is, the more weight past observations have.

\subsection{Mean Hybrid GARCH model}\label{Section:MGARCHLSTM}
The estimated parameters of the above three models (GARCH, EGARCH, EWMA) will be fed into an LSTM to enrich the information, further improving the prediction accuracy of our model. First, we present the first prediction method, namely the Mean Hybrid GARCH (M-H-GARCH) model, based on the work of \cite{kim2018forecasting} as follows, 
\begin{equation}\label{Eq:MHGARCH}
    \begin{split}
& Y_t=\mu_t+\varepsilon_t, \\
& \varepsilon_t=\sigma_t \eta_t, \varepsilon_t \mid \mathcal{F}_{t-1} \sim N\left(0, \sigma_t^2\right),\\
&\sigma_t=G(\Theta,\mathcal{F}_{t-1}),
    \end{split}
\end{equation}
where $Y_t$ is the financial series, e.g., log-returns; $\eta_t \overset{i.i.d.}{\sim} N\left(0,1\right)$; $\mu_t$ is the mean part of $Y_t$, which is approximated by the mean of historical observations $\hat{\mu}_t := \frac{1}{p}\sum^p_{i=1} Y_{t-i}$, where $p$ determines the number of previous observations to be averaged; $\varepsilon_t := \sigma_t\cdot\eta_t$ represents the error term from the mean at a given time $t$; $\Theta$ represents all parameters from GARCH, EGARCH and EWMA models; $G(\cdot)$ is the LSTM model which takes the hidden information implied by parameter set $\Theta$ and other explanatory variables generated from $\mathcal{F}_{t-1}$ as input and returns an estimation of the conditional variance as its output. 

Notably, \cite{hansen2005forecast} provided compelling
evidence that the GARCH(1,1) model cannot be beaten by variants with other order structures. Moreover, the GARCH(1,1) model is the parsimonious version of the ARCH($\infty$) model. Therefore, for simplicity, we consider the GARCH(1,1) model, which is the most popular GARCH-type model to fit the volatility of financial time series, in our hybrid modeling. In addition to the GARCH(1,1) model, the hybrid model also incorporates the estimated parameters from EGARCH(1,1) and EWMA models, since they have different advantages in modeling volatility; see \cite{kim2018forecasting} for more discussion on this point. In short, the hybrid model attempts to leverage all the high-level information implied by the parameters of GARCH, EGARCH, and EWMA models. As a result, we define $\Theta := \bigl(\alpha_0^{(G)}, \alpha_1^{(G)}, \beta_1^{(G)}, \alpha_0^{(E)}, \alpha_1^{(E)}, \gamma_1^{(E)}, \theta_1^{(E)}, \lambda^{(W)}\bigr)^{\top}$, where the superscripts $G$, $E$ and $W$ refer to the GARCH(1,1), EGARCH(1,1) and EWMA processes, respectively, as defined in Sections~\ref{Subsec:GARCH} to~\ref{Subsec:EWMA}. For the LSTM model, we utilize the basic three-gate structure with an input gate, a forget gate and an output gate. Our LSTM is constructed as
\begin{equation}\label{Eq:LSTM}
\begin{split}
g_t &= f^{s}(U_g x_t + W_g h_{t-1} + b_g),\\
i_t &= f^{s}(U_i x_t + W_i h_{t-1} + b_i),\\
\tilde{c}_t &= \tanh(U_c x_t + W_c h_{t-1} + b_c),\\
c_t &= g_t \odot c_{t-1} + i_t \odot \tilde{c}_t,\\
o_t &= f^{s}(U_o x_t + W_o h_{t-1} + b_o),\\
h_t &= o_t \odot \tanh(c_t),
\end{split}
\end{equation}
where $i_t$, $g_t$ and $o_t$ denote the input, forget and output gates, respectively; $c_t$ is the memory cell and $h_t$ the hidden state of the LSTM at time $t$; and the input $x_t$ is the time-indexed vector $\Theta_t = \bigl(\alpha_{0,t}^{(G)}, \alpha_{1,t}^{(G)}, \beta_{1,t}^{(G)}, \alpha_{0,t}^{(E)}, \alpha_{1,t}^{(E)}, \gamma_{1,t}^{(E)}, \theta_{1,t}^{(E)}, \lambda_t^{(W)}\bigr)^{\top}$. In the equations above, $U_\cdot$ and $W_\cdot$ are weight matrices and $b_\cdot$ are bias vectors for $\cdot \in \{g, i, c, o\}$; $f^{s}(z) = 1/(1+e^{-z})$ is the sigmoid activation function; $f^{s}(\cdot)$ and $\tanh(\cdot)$ are applied element-wise; and $\odot$ denotes the element-wise product.

To estimate the LSTM that is deployed to approximate the conditional variance $\sigma_t^2$, we use mean squared error (MSE) as the loss criterion. However, the true $\sigma_t^2$ is never observed in practice. We only observe the financial time series $\{Y_t\}$. To incorporate the model setting \eqref{Eq:MHGARCH}, we define our MSE loss criterion to train LSTM as follows:
\begin{equation}\label{Eq:MSEloss}
 MSE = \frac{1}{T}\sum^T_{t=1}(\hat{\sigma}_t -(\sqrt{(\pi/2)}\cdot|{\hat{\varepsilon}_t}|))^2,  
\end{equation}
where $T$ is the total number of predictions; $\hat{\sigma}_t$ is the predicted volatility at time $t$ by our M-H-GARCH model; $\hat{\varepsilon}_t := Y_t-\hat{\mu}_t$ such that $\hat{\varepsilon}^2_t$ is an asymptotically unbiased estimator of $\sigma_t^2$ provided $\hat{\mu}_t$ converges to the true $\mu_t$ in probability under the ergodic condition of the financial time series $\{Y_t\}$. 

To understand the form of the loss criterion in \eqref{Eq:MSEloss}, we start with a simple fact, i.e., if $Z\sim N(0,1)$, then the expectation of $|Z|$ is $\mathbb{E}(|Z|) = \sqrt{\frac{2}{\pi}}$. Moreover, by the model setting $\varepsilon_t=\sigma_t \eta_t$ is assumed to be $N\left(0, \sigma_t^2\right)$ given historical information $\mathcal{F}_{t-1}$. Thus, $\mathbb{E}\left(\left|\varepsilon_t\right| \big| \mathcal{F}_{t-1}\right) = \sigma_t \sqrt{\frac{2}{\pi}}$. Equivalently, $\mathbb{E}\left[\sqrt{\frac{\pi}{2}}\left|\varepsilon_t\right| \big| \mathcal{F}_{t-1}\right]=\sigma_t$. In other words, $\sqrt{\frac{\pi}{2}}\left|\varepsilon_t\right|$ is an unbiased proxy for the latent conditional standard deviation $\sigma_t$. Since $\sqrt{\frac{\pi}{2}}\left|\hat{\varepsilon}_t\right|$ converges to $\sqrt{\frac{\pi}{2}}\left|\varepsilon_t\right|$ asymptotically with a consistent model estimation, we take the loss criterion \eqref{Eq:MSEloss} to train the LSTM empirically so that it can estimate the conditional variance.


\subsection{ARMA Hybrid GARCH model}\label{Subsec:ARMA-H-GARCH}

Instead of modeling the mean part of the financial series $Y_t$ by $\mu_t$, which is simply the average value of the previous values, we further propose the ARMA-Hybrid-GARCH (ARMA-H-GARCH) model:
\begin{equation}\label{Eq:ARMAHGARCH}
    \begin{split}
& Y_t= c + \sum_{l=1}^L \phi_{l}\cdot Y_{t-l}+\sum_{k=1}^K \psi_{k} \varepsilon_{t-k}
+\varepsilon_t, \\
& \varepsilon_t=\sigma_t \eta_t, \varepsilon_t \mid \mathcal{F}_{t-1} \sim N\left(0, \sigma_t^2\right),\\
&\sigma_t=G(\Theta,\mathcal{F}_{t-1}).
    \end{split}
\end{equation}
One difference between model setting \eqref{Eq:MHGARCH} and \eqref{Eq:ARMAHGARCH} hinges on the mean part of $Y_t$ series. The ARMA-H-GARCH model takes advantage of the ARMA model to regularize the mean part of the financial series. Besides, unlike the M-H-GARCH model, in which the parameters of the mean and variance parts are estimated separately by the sample mean and the LSTM, we propose a maximum likelihood estimation algorithm to estimate all parameters in the ARMA-H-GARCH model simultaneously.

More specifically, let $\Pi=\left(c, \phi_1, \ldots, \phi_L, \psi_1, \ldots, \psi_K\right)$
denote the ARMA parameters. Our hybrid model is
\begin{equation}\label{Eq:simplifiedAMAH}
    Y_t=\mu_t(\Pi)+\varepsilon_t,
\end{equation}
where $\mu_t(\Pi) := c + \sum_{l=1}^L \phi_l Y_{t-l}+\sum_{k=1}^K \psi_k \varepsilon_{t-k}$ and $
\varepsilon_t \mid \mathcal{F}_{t-1} \sim N\left(0, \sigma_t^2\right)$, $\sigma_t^2=G(\Theta,\mathcal{F}_{t-1})$. Moreover, $\mu_t(\Pi)$ is actually the conditional mean of $Y_t$ given the historical information $\mathcal{F}_{t-1}$, i.e., $\mu_t(\Pi)=E\left(Y_t \mid \mathcal{F}_{t-1}\right)$. Thus, under the Gaussian distribution assumption, we can get the conditional density of $Y_t$ given $\mathcal{F}_{t-1}$ as follows:
\begin{equation*}
    f\left(Y_t \mid \mathcal{F}_{t-1} ; \Pi, \Theta\right)=\frac{1}{\sqrt{2 \pi \sigma_t^2(\Pi, \Theta)}} \exp \left(-\frac{\left[Y_t-\mu_t(\Pi)\right]^2}{2 \sigma_t^2(\Pi, \Theta)}\right),
\end{equation*}
where $\sigma_t^2(\Pi, \Theta)$ is the volatility at the time index $t$ which depends on all model parameters. 
By the chain rule of conditional probability, we have that the joint density of $\{Y_{t_0},\ldots, Y_{T}\}$ is equivalent to 
\begin{equation}\label{Eq:conditionalMLE}
   f\left(Y_{t_0},\ldots,Y_{T}  \mid \mathcal{F}_{t_0 -1} ; \Pi, \Theta\right) = \prod^T_{t=t_0}f\left(Y_t  \mid \mathcal{F}_{t-1} ; \Pi, \Theta\right),
\end{equation}
where $t_0=\max (L, K)+1$ and we treat the initial values $\{Y_1,\ldots,Y_{t_0-1}\}$ as fixed. More specifically, the conditional likelihood is applied here.  Taking the logarithm of both sides of \eqref{Eq:conditionalMLE} and dropping the constant irrelevant terms, the negative log-likelihood is
\begin{equation}\label{Eq:LLF}
    \mathcal{L}(\Pi, \Theta)=\frac{1}{2 N} \sum_{t=t_0}^T\left[\log \sigma_t^2(\Pi, \Theta)+\frac{\varepsilon_t(\Pi)^2}{\sigma_t^2(\Pi, \Theta)}\right],
\end{equation}
where $N := T-t_0 + 1$; $\varepsilon_t := \varepsilon_t(\Pi) = Y_t-\mu_t(\Pi)$ based on the ARMA structure of the mean part, where the $\varepsilon_t$ can be seen as the residual, which also depends on $\Pi$. Subsequently, the LSTM model and the ARMA model can be trained together by minimizing the negative log-likelihood function (\ref{Eq:LLF}). The training procedure can be achieved by a specific gradient descent algorithm easily.

\section{Bootstrap prediction interval}\label{Sec: BootPI}
In this section, we introduce the algorithms to build the bootstrap prediction interval with M-H-GARCH and ARMA-H-GARCH models. To implement the algorithms, we introduce two approaches based on normal errors and the empirical distribution of residuals, respectively. Then, we propose a systematic way inspired by conformal prediction to widen the interval length so that the notorious undercoverage issue of bootstrap prediction intervals can be mitigated. 
\subsection{M-H-GARCH approach}
We first give the forward bootstrap prediction interval of the M-H-GARCH model. Suppose we have obtained the trained LSTM model, $\widehat{G}(\cdot)$; and $\hat{\mu}_{T}$, which is the estimated conditional mean of $Y_t$ over a window of $p$. In practice, the value of $p$ is a tuning parameter that can be selected by cross-validation. Our goal is to make point predictions and prediction intervals of future realized volatility, which is defined as follows:
\begin{equation}\label{Eq:futureVolatility}
    R V_{T,H}=\sqrt{\frac{1}{H} \sum_{h=1}^{H}\left(Y_{T+h}-\overline{Y}_{T,H}\right)^2};
\end{equation}
where $\overline{Y}_{T,H}$ is the average of the log-returns from time $T+1$ to $T+H$ and $H$ is the forecast horizon (number of future trading days). The detailed algorithm to make point predictions and prediction intervals according to the M-H-GARCH approach is illustrated in  Algorithm \ref{Algo:PIMHGARCH}. The hyperparameters for the training procedure are given in Appendix \ref{sec:implementation}.

\begin{algorithm}
\caption{The bootstrap prediction interval for M-H-GARCH model}\label{Algo:PIMHGARCH}
\setstretch{1.25}
\begin{algorithmic}[1]
\Require the initial LSTM Model $G(\cdot)$; the observed data series $\{Y_1, Y_2, ..., Y_T\}$; the hyperparameters for training.
\State  Get the estimated LSTM model $\widehat{G}(\cdot)$.
\For{$h$ in $1:H$} 
\State Generate $\eta^*_{T+h}$ from $F_\eta = N(0,1)$ and compute $\varepsilon_{T+h}^* = \hat{\sigma}_{T+h} \cdot \eta^*_{T+h}$, where $\hat{\sigma}_{T+h}$ is $\widehat{G}(\widehat{\Theta},\mathcal{F}_{T+h-1})$.  
\State Derive the estimator $\hat{\mu}_{T+h}$ by $\frac{1}
{p}\sum_{i=1}^p Y_{T+h-i}$.
\State Compute $Y^*_{T+h} := \hat{\mu}_{T+h} + \varepsilon_{T+h}^*$.
\State Take $Y_{T+h}$ as $Y^*_{T+h}$ and then redefine the corresponding $\mathcal{F}_{T+h}$.
\EndFor
\State Repeat the above for loop $M$ times, so we can get $M$ number of $H$ step ahead pseudo series. Denote them as $\{Y^{(m)}_{T+1}, \ldots, Y^{(m)}_{T+H} \}_{m=1}^M$. 
\State Compute realized volatility based on $M$ number of $H$ step ahead pseudo series. As a result, we can obtain $\{\widehat{RV}^{(m)}_{T,H}\}_{m=1}^M$. 
\State Determine the empirical distribution of $\{\widehat{RV}^{(m)}_{T,H}\}_{m=1}^M$ as $\widehat{F}_{RV}$. 
\State \textbf{Return} The optimal $L_2$ prediction of $H$-step ahead realized volatility as the mean of $\widehat{F}_{RV}$. The corresponding $(1-\alpha)$-PI as the $[\hat{q}_{\alpha/2}, \hat{q}_{1-\alpha/2}]$, where $\hat{q}_{\alpha/2}$ and $\hat{q}_{1-\alpha/2}$ are $\alpha/2$ and $1-\alpha/2$ lower quantiles of $\widehat{F}_{RV}$.
\end{algorithmic}
\end{algorithm}

\FloatBarrier
\subsection{ARMA-H-GARCH approach}
Rather than approximating the mean part of the time series by the sample mean over a latest information window, the ARMA-H-GARCH model estimates the conditional mean of $Y_t$ given historical information by an ARMA model. As illustrated in Section \ref{Subsec:ARMA-H-GARCH}, all parameters involved in the ARMA-H-GARCH model can be estimated by maximizing the joint likelihood function. Similar to  Algorithm \ref{Algo:PIMHGARCH}, we propose our  Algorithm \ref{Algo:PIARMAHGARCH} to make point predictions and prediction intervals in the ARMA-H-GARCH approach.

\begin{algorithm}
\caption{The bootstrap prediction interval for ARMA-H-GARCH model}\label{Algo:PIARMAHGARCH}
\setstretch{1.25}
\begin{algorithmic}[1]
\Require the initial LSTM Model $G(\cdot)$; the observed data series $\{Y_1, Y_2, ..., Y_T\}$; the hyperparameters for training.
\State  Get estimated LSTM model $\widehat{G}(\cdot)$ and ARMA model parameters $\hat\alpha$ jointly, through minimizing the likelihood \eqref{Eq:LLF}.
\For{$h$ in $1:H$} 
\State Generate $\eta^*_{T+h}$ from $F_\eta \sim N(0,1)$ and compute $\varepsilon_{T+h}^* = \hat{\sigma}_{T+h} \cdot \eta^*_{T+h}$, where $\hat{\sigma}_{T+h}$ is $\widehat{G}(\widehat{\Theta},\mathcal{F}_{T+h-1})$. 
\State Derive the estimator $\hat{\mu}_{T+h}= \hat{c} + \sum_{l=1}^L \hat\phi_{l}\cdot Y_{T+h-l}+\sum_{k=1}^K \hat\psi_{k}\cdot \varepsilon^*_{T+h-k}$. For $k\geq h$, $\varepsilon^*_{T+h-k}$ is taken as the estimated $\hat{\varepsilon}_{T+h-k}$ under the ARMA-H-GARCH setting, i.e., the in-sample fit.
\State Compute $Y^*_{T+h} := \hat{\mu}_{T+h} + \varepsilon_{T+h}^*$.
\State Take $Y_{T+h}$ as $Y^*_{T+h}$ and then redefine the corresponding $\mathcal{F}_{T+h}$.
\EndFor
\State Repeat the above for loops similar to steps 8 to 11 in  Algorithm \ref{Algo:PIMHGARCH} to determine $L_2$ point prediction of $ H$-step-ahead realized volatility and build the corresponding prediction interval.
\end{algorithmic}
\end{algorithm}

\begin{Remark}[Make the empirical distribution of $\eta$]\label{Remark:empiricaleta}
    In our model setting (\ref{Eq:MHGARCH}) and (\ref{Eq:ARMAHGARCH}), we assume $\eta_t\sim N\left(0, 1\right)$. However, it is well known that the distribution of financial series typically possesses a fatter tail than normal. As a result, the variance of the future volatility may be underestimated if $\eta^*_{T+h}$ are generated from $N(0,1)$. Here, we provide an alternative approach to generate $\eta^*_{T+h}$ from its own empirical distribution. In particular, 
    \begin{itemize}
        \item For the M-H-GARCH model, we can define $\widehat{F}_{\eta}$ as the empirical distribution of $\{\frac{Y_1 - \hat{\mu}_1}{\hat{\sigma}_1} -\bar{c},\ldots,$ $ \frac{Y_T - \hat{\mu}_T}{\hat{\sigma}_T} -\bar{c}\}$, where $\{\hat{\mu}_t\}$ and $\{\hat{\sigma}_t\}$ are estimated mean and variance at time index $t$, respectively, for $t=1,\ldots,T$; $\bar{c}$ is the sample mean of $\{\frac{Y_1 - \hat{\mu}_1}{\hat{\sigma}_1},\ldots,$ $ \frac{Y_T - \hat{\mu}_T}{\hat{\sigma}_T}\}$. More specifically, $\hat{\mu}_t$ and $\hat{\sigma}_t$ are estimated by a rolling mean and an LSTM, respectively. To simplify the notation, we assume the initial observations before time index 1 exist. 
        \item For the ARMA-H-GARCH model, we can define $\widehat{F}_{\eta}$ as the empirical distribution of $\{\frac{Y_1 - \hat{\mu}_1(\hat{\Pi})}{\hat{\sigma}_1} - \bar{c},\ldots,$ $ \frac{Y_T - \hat{\mu}_T(\hat{\Pi})}{\hat{\sigma}_T} - \bar{c} \}$, where $\{\hat{\mu}_t(\hat{\Pi})\}$ and $\{\hat{\sigma}_t\}$ are estimated mean and variance at time index $t$, respectively, for $t=1,\ldots,T$; $\bar{c}$ is the sample mean of $\{\frac{Y_1 - \hat{\mu}_1(\hat{\Pi})}{\hat{\sigma}_1},\ldots,$ $ \frac{Y_T - \hat{\mu}_T(\hat{\Pi})}{\hat{\sigma}_T}\}$. More specifically, $\hat{\mu}_t(\hat{\Pi})$ is the mean function based on the ARMA structure defined in (\ref{Eq:simplifiedAMAH}) with estimated parameter vector $\hat{\Pi}$. To simplify the notation, we still assume the initial observations before time index 1 exist. 
    \end{itemize}

\end{Remark}

Then, we can get two variants of the Algorithms \ref{Algo:PIMHGARCH} and \ref{Algo:PIARMAHGARCH} by generating $\eta^*_{T+h}$ from $\widehat{F}_{\eta}$ instead of $N(0,1)$. The performance of these variants will be compared in Section \ref{Sec:simulation}. Moreover, a systematic way to extend our prediction interval to improve the empirical coverage rate will be provided in Section \ref{Sec:conformalexten}. Lastly, we should mention that the estimation procedure of ARMA-H-GARCH is still based on the likelihood function from normal. Thus, the application of $\widehat{F}_{\eta}$ stands for a way to mimic the fat tail of financial series distribution.

\subsection{Conformal extension on interval}\label{Sec:conformalexten}
It is well known that prediction intervals suffer from undercoverage in practice, especially in finite-sample settings. In particular, \cite{politis2015model} provided a perspective to understand this undercoverage issue, i.e., the standard prediction interval does not capture the model's estimation variability; see the following Remark \ref{Remark:effectsofmodelestimation} to understand the effects of model estimation variability on influencing the coverage rate of prediction intervals.

\begin{Remark}\label{Remark:effectsofmodelestimation}
    Suppose we consider the linear regression model $Y_i = X_i^T\beta + \epsilon_i$ with $n$ samples. By the least squares technique, we can get the estimation of the parameter coefficient $\hat{\beta}$. Given a new predictor $x_f$, we have the predictive root
$
Y_f - \widehat{Y}_f := x_f^T\beta + \epsilon_f - x_f^T\hat{\beta} = \epsilon_f + e_f,
$
where $e_f := x_f^T\beta -  x_f^T\hat{\beta}$ stands for the estimation error. Under standard conditions and a well-specified model,  $
E(x_f^T\hat{\beta}) = x_f^T\beta~\text{and}~Var(x_f^T\hat{\beta}) = O(1/n),$
as $n\to\infty$. Since $e_f\overset{p}{\to}0$, it is tempting to apply 
\begin{equation}\label{Eq:naivePIlinearReg}
    [\widehat{Y}_f + \widehat{F}^{-1}_{\epsilon,\alpha/2}, \widehat{Y}_f + \widehat{F}^{-1}_{\epsilon,1-\alpha/2}]
\end{equation}
as a PI, where $\widehat{F}^{-1}_{\epsilon,\alpha/2}$ and $\widehat{F}^{-1}_{\epsilon,1-\alpha/2}$ are $\alpha/2$ and $1-\alpha/2$ quantiles of $\widehat{F}_{\epsilon}$, which is the empirical distribution of residuals $\{r_i\}_{i=1}^n$ from regression. In other words, PI (\ref{Eq:naivePIlinearReg}) treats $e_f$ as 0 and typically results in undercoverage. However, by Chebyshev's inequality, $e_f = O_p(1/\sqrt{n})$, which is not negligible with $n$ being small. Thus, it is important to take the model estimation variability into account when the prediction interval is built. 

\end{Remark}

There are multiple ways to improve the coverage rate of prediction intervals. For example, \cite{politis2015model} defined a so-called pertinent prediction interval (PPI) to capture the model estimation variability by approximating the distribution of the predictive root with the bootstrap technique; see more explanation and applications of this pertinent prediction interval in \cite{wang2026model, wu2024deep,wu2026prediction}. However, the procedure to build the PPI requires the repetition of the training procedure with the bootstrap data multiple times, which is computationally heavy when the complex model estimator is adopted. On the other hand, conformal prediction, which can guarantee the marginal coverage, introduced by \cite{vovk2005algorithmic}, has become increasingly popular recently; see variants of conformal prediction methods from \cite{chernozhukov2021distributional,sesia2021conformal,romano2019conformalized,lei2018distribution}, for other examples. Besides, we should mention that the PPI and conformal prediction can be related; see that one specific type of PPI shares a common intermediate step of making the special conformal prediction in \cite{chernozhukov2021distributional,dai2026distributional}.

In our case, since the complex LSTM is used for our prediction problems, we propose a systematic way inspired by the conformal prediction idea to extend the length of the prediction intervals returned by  Algorithms \ref{Algo:PIMHGARCH} and \ref{Algo:PIARMAHGARCH}. In short, we use a recursive prediction scheme to compute the conformal score based on the miscoverage error in the recursive prediction interval. Then, the lower bound and upper bound of the targeted prediction interval are adjusted with the conformal score. The concrete algorithm to perform this so-called conformal extension for our prediction interval is provided in  Algorithm \ref{Algo:conformalextension}.

\begin{algorithm}
\caption{Conformal extension of prediction intervals returned by Algorithms \ref{Algo:PIMHGARCH} and \ref{Algo:PIARMAHGARCH}}\label{Algo:conformalextension}
\label{Algo:conformalexten}
\setstretch{1.25}
\begin{algorithmic}[1]
\Require The requirements of Algorithm \ref{Algo:PIMHGARCH} or Algorithm \ref{Algo:PIARMAHGARCH}; the initial window size $\varpi$ to compute the conformal score. 
\State Perform the necessary estimations on unknown parameters in Algorithm \ref{Algo:PIMHGARCH} or Algorithm \ref{Algo:PIARMAHGARCH} with all data. 
\State Define $J := T - H - \varpi$ + 1, i.e., the total number of conformal scores.
\For{$j$ in $1:J$} 
\State Take data $\{Y_1,\ldots,Y_{{\varpi}+j-1}\}$ to build a prediction interval for $R V_{\varpi+j-1,H}$ according to either Algorithm \ref{Algo:PIMHGARCH} or Algorithm \ref{Algo:PIARMAHGARCH} with model estimation from Step 1. Denote the lower and upper bounds of the prediction interval as $L_{\varpi+j-1}^r$ and $U_{\varpi+j-1}^r$, respectively. The superscript ``r'' means the recursive prediction scheme is taken.
\State Compute the conformal score at the $(\varpi+j-1)$-th recursive prediction as 
\begin{equation}\label{Eq:conformalscore}
    S_{\varpi+j-1} :=\frac{\max \left\{L_{\varpi+j-1}^r-\widehat{RV}_{{\varpi+j-1}, H}, \widehat{RV}_{{\varpi+j-1}, H}-U_{\varpi+j-1}^r, 0\right\}}{U_{\varpi+j-1}^r - L_{\varpi+j-1}^r},
\end{equation}
where $\widehat{RV}_{{\varpi+j-1}, H}$ is computed from the realized time series. 
\EndFor
\State Get $J$ conformal scores $\{S_{\varpi}, \ldots, S_{T-H}\}$.
\State Apply Algorithm \ref{Algo:PIMHGARCH} or Algorithm \ref{Algo:PIARMAHGARCH} to build the prediction interval for $R V_{T,H}$ with data $\{Y_1, Y_2, \ldots, Y_T\}$. Denote the corresponding lower and upper bounds as $L_{T}$ and $U_{T}$, respectively. 
\State Make the conformal extension prediction interval as
\begin{equation}\label{Eq:conformalextenPI}
\left[L_T-\widehat{q}_{S,1-\alpha} W_T, U_T+\widehat{q}_{S,1-\alpha} W_T\right],
\end{equation}
where $W_T := U_T - L_T$ and $\widehat{q}_{S,1-\alpha}$ is the $1-\alpha$ empirical quantile of $\{S_{\varpi}, \ldots, S_{T-H}\}$.
\end{algorithmic}
\end{algorithm}

Since the conformal score defined in (\ref{Eq:conformalscore}) of Algorithm \ref{Algo:conformalextension} is always non-negative, the prediction interval (\ref{Eq:conformalextenPI}) is always at least as wide as the initial prediction interval returned by Algorithm \ref{Algo:PIMHGARCH} or Algorithm \ref{Algo:PIARMAHGARCH}. Thus, Algorithm \ref{Algo:conformalexten} stands for a systematic way to extend the prediction interval. In the simulation studies, we use the window size $\varpi = 200$ for $T = 250$ and $\varpi = 400$ for $T = 500$. In addition, since $RV_{T,H}\geq 0$, we truncate the conformal extension prediction interval from below at zero.
However, we should mention that our prediction interval (\ref{Eq:conformalextenPI}) does not have the typical marginal coverage guarantee property of the conformal prediction interval since the exchangeability of conformal scores $\{S_{\varpi}, \ldots, S_{T-H}\}$ is not satisfied. In fact, these conformal scores are correlated to each other since the calculation of $\widehat{RV}_{{\varpi+j-1}, H}$ for $j = 1,\ldots, J$ involves overlapping of observed series. Also, the estimated model used to do recursive forecasting is trained with the whole data set. A stricter way to compute the conformal score is to separate each conformal score by $H$ time index, i.e., we only compute $\{S_{\varpi+(k-1)H}\}$ for $k = 1,\ldots, \left\lfloor\frac{T-\varpi}{H}\right\rfloor$. In other words, the dependence between the adjacent conformal scores is partially dampened by the separation. Moreover, we should retrain the model only with data from the recursively expanding window. However, this refitting procedure unavoidably increases the computational burden.

Furthermore, if $\{S_{\varpi+(k-1)H}, S_{\varpi+H},\ldots, S_{T-H}\}$ are strictly stationary and ergodic under necessary conditions for time series and model estimator, it is possible to show that the prediction interval (\ref{Eq:conformalextenPI}) satisfies the marginal coverage guarantee asymptotically, i.e., 
$$\mathbb{P}(RV_{T,H}\in \left[L_T-\widehat{q}_{S,1-\alpha} W_T, U_T+\widehat{q}_{S,1-\alpha} W_T\right]) \to 1-\alpha,$$ 
as $T\to\infty$, $\varpi \to \infty$ and $\left\lfloor\frac{T-\varpi}{H}\right\rfloor \to \infty$. We decide to postpone the theoretical analysis of the conformal extension prediction interval to future work. In this work, the performance of the prediction interval (\ref{Eq:conformalextenPI}) will be evaluated by simulation and real data studies.
\FloatBarrier
\section{Simulations}\label{Sec:simulation}



For the M-H-GARCH method, the conditional mean is the rolling average over the most recent $p$ observations. In practice, $p$ is a tuning parameter whose value can be determined by cross-validation. In our simulation studies, we take $p = \{1,\ldots,5\}$ to ensure the simulation is comprehensive and report the optimal $p$ according to the root mean squared error of the point predictions.  For the ARMA-H-GARCH method, the orders $(L, K)$ are selected per replication by AIC over $L, K \leq 5$. The chosen order is then estimated using the conditional likelihood of Eq.~\eqref{Eq:LLF} with $t_0 = \max(L, K)+1$. For the naive LSTM method, the mean component is also taken as the rolling average, as in the M-H-GARCH model. This results in the difference between naive LSTM and M-H-GARCH models being due to the inputs of the models; see Appendix \ref{sec:implementation} for more implementation details of different methods.




We assess each method by implementing Monte Carlo replication. For a given data-generating process and sample size $T$, one replication consists of simulating an independent series $\{Y_1,\dots,Y_{T+H}\}$, fitting every method on the first $T$ observations, generating the point prediction, constructing a $95\%$ prediction interval for $RV_{T,H}$ as defined in Eq.~\eqref{Eq:futureVolatility}, and recording whether that interval contains the realized value computed from the held-out continuation $\{Y_{T+1},\dots,Y_{T+H}\}$. Three quantities are held fixed throughout:

\begin{itemize}
  \item $T \in \{250, 500\}$, the length of each simulated series
        (short data mimics one year of econometric data, long data two
        years);
  \item $M = 2000$, the number of bootstrap draws used to form a single prediction interval;
  \item $R = 500$, the number of independent replications over which
        coverage and interval length are averaged.
\end{itemize}
We consider two data-generating processes, which cover $4$ scenerios in combination with the two sample sizes.

\medskip
\noindent\textbf{Model-1} \quad Standard GARCH(1,1) with Student-$t$ errors and ARMA(1,1) mean structure
\begin{align*}
Y_t &= 0.0002 + 0.5\,Y_{t-1} + 0.3\,\varepsilon_{t-1} + \varepsilon_t, \\
\varepsilon_t &= \sigma_t \eta_t, \qquad
\sigma_t^2 = 0.00001 + 0.73\,\sigma_{t-1}^2 + 0.1\,\varepsilon_{t-1}^2,
\end{align*}
with $\{\eta_t\}\sim$ i.i.d.\ Student-$t$ with five degrees of freedom, standardized to unit variance. \\

\noindent\textbf{Model-2} \quad Time-varying GARCH(1,1) with Student-$t$ errors and ARMA(1,1) mean structure
\begin{align*}
Y_t &= 0.0002 + 0.5\,Y_{t-1} + 0.3\,\varepsilon_{t-1} + \varepsilon_t, \\
\varepsilon_t &= \sigma_t\eta_t,\qquad \sigma_t^2 = 0.00001 + \beta_{1,t}\sigma_{t-1}^2+\alpha_{1,t}\varepsilon_{t-1}^2,\\
V_t &= t/T, \qquad \alpha_{1,t} = 0.1 - 0.05V_t; \beta_{1,t} = 0.55 + 0.25V_t,~\text{for}~t = 1,\ldots, T+H.
\end{align*}
with $\{\eta_t\}\sim$ i.i.d.\ Student-$t$ with five degrees of freedom, standardized to unit variance; $T = 250+H$ or $500 + H$ in our simulation cases; $H = 5$ and $H=20$ represent one-week and one-month ahead predictions, respectively. All simulation series are generated with 500 discarded burn-in observations; see Appendix \ref{sec:implementation} for more details on the simulation procedure to guarantee reproducibility.

We use these two models to challenge our methods when the model is misspecified. For Model-1, we take the $t$-distribution to mimic the heavy-tail distribution of the real financial data. For Model-2, we take the time-varying GARCH coefficients to mimic the dynamic situation in the real world. To simulate data from Model-1, we take initial values $Y_0 = 0.0004$ and $\sigma^2_0 = 5.88 \times 10^{-5}$. To simulate data from Model-2, we take initial values $Y_0 = 0.0004$ and $\sigma^2_0 = 2.86 \times 10^{-5}$. These values represent the unconditional mean and variance in Model-1 and Model-2 at $t=0$, respectively. 

In order to calculate the average interval length and coverage, we write $[\,\widehat{L}^{(r)}, \widehat{U}^{(r)}\,]$ for the interval produced on the $r$-th replication of each method and $RV^{(r)}_{T,H}$ for the realized value computed from that replication's held-out continuation. The reported coverage and average interval length are thus
\begin{equation*}
\widehat{C}_{1-\alpha} \;=\; \frac{1}{R}\sum_{r=1}^{R}
  \mathbf{1}\!\left\{ \widehat{L}^{(r)} \le RV^{(r)}_{T,H} \le \widehat{U}^{(r)} \right\},
\qquad
\widehat{W}_{1-\alpha} \;=\; \frac{1}{R}\sum_{r=1}^{R}\left( \widehat{U}^{(r)} - \widehat{L}^{(r)} \right).
\end{equation*}
The RMSE (root mean squared error), MAPE (mean absolute percentage error) and MDAE (median absolute error) of the point prediction are computed from the same $R$ replications in an analogous way. 

Throughout the study, we consider five methods: (1) M-H-GARCH; (2) ARMA-H-GARCH; (3) GARCH(1,1); (4) EGARCH(1,1); and (5) a naive LSTM. GARCH(1,1) and EGARCH(1,1) are determined by maximum likelihood estimation; the EWMA decay is estimated with initial value $\lambda = 0.94$. Both GARCH and EGARCH are estimated by Gaussian quasi-maximum likelihood with the ${\tt arch}$ package in ${\tt Python}$. Prediction intervals based on methods (3)--(5) are constructed with the same forward bootstrap as Algorithms~\ref{Algo:PIMHGARCH} and~\ref{Algo:PIARMAHGARCH}. All forward bootstrap prediction intervals use $M = 2000$ bootstrap draws at the $\alpha = 0.05$ level. For methods (1) and (2), we also provide the prediction intervals extended by conformal prediction and generating future $\eta^{*}_{T+h}$ from $\widehat{F}_\eta$, as defined in Remark \ref{Remark:empiricaleta}. Thus, in total, we have 11 different prediction intervals to compare. For each method, we report (1) the root mean squared error (RMSE) of the point prediction; (2) the mean absolute percentage error (MAPE) of the point prediction; (3) the median absolute error (MDAE) of the point prediction; (4) the empirical coverage rate of the $95\%$ prediction interval; and (5) the interval length. All these metrics are averaged over $R=500$ replications. In addition, the sample standard deviations of metrics (1), (2), (4) and (5) over 500 replications are also provided.

\subsection{Simulation Results}

All simulation results are presented in Tables \ref{Table:M1H5} to \ref{Table:M2H20}, where sample standard deviations over replications are given in parentheses; best values, i.e., smallest for RMSE, MAPE and MDAE, closest to 0.95 for Coverage and shortest interval length are bold. From the results, we summarize several findings as follows:
\begin{itemize}
    \item[1.] For point prediction, the ARMA-H-GARCH method dominates other methods. Generally speaking, it has the smallest RMSE, MAPE and MDAE for the two simulation models. Moreover, it always outperforms the naive LSTM method on point prediction. On the other hand, the M-H-GARCH method based on \cite{kim2018forecasting} does not guarantee consistently superior performance over the naive LSTM method in point predictions. In short, the implication is two-fold:  (1) using the estimates of variance and the parameters from classical models improves the accuracy of the LSTM's realized volatility forecast;  (2) using bootstrap prediction intervals, the ARMA mean structure is superior to a rolling mean estimate when estimating the mean of financial series. We also note that the naive LSTM method is beaten by GARCH(1,1) and EGARCH(1,1). This implies that the solely network-based method does not outperform the classical methods in some situations. 

    \item[2.] For the prediction intervals, all optimal coverage rates come from our ARMA-H-GARCH or M-H-GARCH method. Moreover, all optimal coverage rates are obtained based on generating $\eta_{T+h}^*$ from its empirical distribution $\widehat{F}_{\eta}$, which emphasizes the importance of mimicking the fat tail property of financial series as we mentioned in Remark \ref{Remark:empiricaleta}. In addition, the conformal extension always makes the coverage rate larger than the variant without this modification, though the sacrifice is the wider prediction interval. 
\end{itemize}




\begin{table}[ht]\centering\scriptsize
\caption{Simulation of Model-1 with H = 5.}
\label{Table:M1H5}
\resizebox{\textwidth}{!}{%
\begin{tabular}{llccccc}\toprule
Method & $\eta_{T+h}^*$ & RMSE($\times 10^{-3}$) & MAPE(\%) & MDAE($\times 10^{-3}$) & Coverage & Length($\times 10^{-3}$) \\ \midrule
\multicolumn{7}{c}{$T = 250$} \\ [8pt]
M-H-GARCH ($p=5$) & $N(0,1)$ & 4.78 (3.22) & 86.60 (104.87) & 3.00 & 0.844 (0.363) & 12.45 (5.22) \\
 & $\hat F_\eta$ &  & &  & 0.850 (0.357) & 13.61 (5.89) \\
+ conformal & $N(0,1)$  &  & &  & 0.900 (0.300) & 14.49 (9.56) \\
+ conformal & $\hat F_\eta$  &  & &  & 0.906 (0.292) & 15.21 (9.56) \\[2pt]
ARMA-H-GARCH & $N(0,1)$ & \textbf{3.79} (2.59) & \textbf{64.15} (77.83) & \textbf{2.17} & 0.884 (0.321) & \textbf{10.66} (2.36) \\
 & $\hat F_\eta$ &  & &  & 0.914 (0.281) & 11.66 (2.90) \\
+ conformal & $N(0,1)$  &  & &  & 0.922 (0.268) & 12.29 (4.52) \\
+ conformal & $\hat F_\eta$  &  & &  & \textbf{0.930} (0.255) & 12.81 (4.57) \\[3pt]
GARCH(1,1) & $N(0,1)$ & 4.78 (3.21) & 90.83 (107.92) & 3.22 & 0.904 (0.295) & 16.84 (8.25) \\[3pt]
EGARCH(1,1) & $N(0,1)$ & 4.44 (2.76) & 89.55 (103.53) & 3.16 & 0.910 (0.286) & 15.68 (4.42) \\[3pt]
naive LSTM ($p=1$) & $N(0,1)$ & 4.39 (3.09) & 73.52 (91.24) & 2.46 & 0.922 (0.268) & 14.36 (7.32) \\[5pt]
\multicolumn{7}{c}{$T = 500$} \\ [8pt]
M-H-GARCH ($p=1$) & $N(0,1)$ & 3.62 (2.19) & 72.08 (84.27) & 2.60 & 0.934 (0.249) & 15.21 (4.79) \\
 & $\hat F_\eta$ &  & &  & \textbf{0.950} (0.218) & 17.11 (5.10) \\
+ conformal & $N(0,1)$  &  & &  & 0.946 (0.226) & 15.64 (4.87) \\
+ conformal & $\hat F_\eta$  &  & &  & 0.954 (0.210) & 17.31 (5.14) \\[3pt]
ARMA-H-GARCH & $N(0,1)$ & \textbf{3.32} (2.12) & \textbf{61.37} (71.31) & \textbf{2.20} & 0.902 (0.298) & \textbf{10.75} (2.32) \\
 & $\hat F_\eta$ &  & &  & 0.924 (0.265) & 11.89 (2.50) \\
+ conformal & $N(0,1)$  &  & &  & 0.946 (0.226) & 12.35 (3.51) \\
+ conformal & $\hat F_\eta$  &  & &  & 0.948 (0.222) & 12.87 (3.34) \\[3pt]
GARCH(1,1) & $N(0,1)$ & 3.89 (2.16) & 88.08 (103.35) & 2.98 & 0.900 (0.300) & 16.29 (4.84) \\[3pt]
EGARCH(1,1) & $N(0,1)$ & 3.87 (2.11) & 88.65 (102.82) & 3.11 & 0.900 (0.300) & 15.53 (2.96) \\[3pt]
naive LSTM ($p=1$) & $N(0,1)$ & 3.72 (2.31) & 73.48 (92.41) & 2.54 & 0.928 (0.259) & 14.93 (5.98) \\
\bottomrule
\end{tabular}}
\end{table}

\begin{table}[ht]\centering\scriptsize
\caption{Simulation of Model-2 with H = 5.}\label{Table:M2H5}
\resizebox{\textwidth}{!}{%
\begin{tabular}{llccccc}\toprule
Method & $\eta_{T+h}^*$ & RMSE($\times 10^{-3}$) & MAPE(\%) & MDAE($\times 10^{-3}$) & Coverage & Length($\times 10^{-3}$) \\  \midrule
\multicolumn{7}{c}{$T = 250$} \\ [8pt]
M-H-GARCH ($p=1$) & $N(0,1)$ & 3.77 (2.63) & 56.50 (70.68) & 2.08 & 0.926 (0.262) & 13.15 (5.33) \\
 & $\hat F_\eta$ &  & &  & \textbf{0.950} (0.218) & 14.71 (5.62) \\
+ conformal & $N(0,1)$  &  & &  & 0.932 (0.252) & 13.76 (5.73) \\
+ conformal & $\hat F_\eta$  &  & &  & 0.954 (0.210) & 14.92 (5.73) \\[2pt]
ARMA-H-GARCH & $N(0,1)$ & \textbf{3.51} (2.52) & \textbf{47.84} (59.48) & \textbf{1.84} & 0.884 (0.321) & \textbf{9.22} (1.81) \\
 & $\hat F_\eta$ &  & &  & 0.904 (0.295) & 10.09 (2.10) \\
+ conformal & $N(0,1)$  &  & &  & 0.916 (0.278) & 10.98 (3.74) \\
+ conformal & $\hat F_\eta$  &  & &  & 0.924 (0.265) & 11.22 (3.35) \\[3pt]
GARCH(1,1) & $N(0,1)$ & 3.73 (2.54) & 63.19 (85.40) & 2.15 & 0.934 (0.249) & 14.17 (5.46) \\[3pt]
EGARCH(1,1) & $N(0,1)$ & 3.71 (2.51) & 63.36 (86.27) & 2.28 & 0.934 (0.249) & 13.15 (3.19) \\[3pt]
naive LSTM ($p=5$) & $N(0,1)$ & 3.67 (2.54) & 57.85 (82.76) & 2.06 & 0.872 (0.334) & 10.23 (4.46) \\[5pt]
\multicolumn{7}{c}{$T = 500$} \\ [8pt]
M-H-GARCH ($p=1$) & $N(0,1)$ & 4.01 (2.80) & 54.54 (66.47) & 2.07 & 0.924 (0.265) & 13.92 (5.20) \\
 & $\hat F_\eta$ &  & &  & 0.944 (0.230) & 15.89 (5.82) \\
+ conformal & $N(0,1)$  &  & &  & 0.930 (0.255) & 14.63 (6.06) \\
+ conformal & $\hat F_\eta$  &  & &  & \textbf{0.950} (0.218) & 16.23 (6.49) \\[3pt]
ARMA-H-GARCH & $N(0,1)$ & 3.90 (2.75) & \textbf{47.98} (54.04) & \textbf{2.06} & 0.864 (0.343) & \textbf{9.61} (1.85) \\
 & $\hat F_\eta$ &  & &  & 0.890 (0.313) & 10.60 (1.95) \\
+ conformal & $N(0,1)$  &  & &  & 0.900 (0.300) & 12.02 (4.20) \\
+ conformal & $\hat F_\eta$  &  & &  & 0.908 (0.289) & 12.17 (3.90) \\[3pt]
GARCH(1,1) & $N(0,1)$ & 3.81 (2.49) & 59.86 (71.92) & 2.41 & 0.936 (0.245) & 14.89 (5.54) \\[3pt]
EGARCH(1,1) & $N(0,1)$ & \textbf{3.74} (2.41) & 58.70 (69.65) & 2.39 & 0.936 (0.245) & 13.54 (2.84) \\[3pt]
naive LSTM ($p=1$) & $N(0,1)$ & 4.07 (2.83) & 53.65 (66.52) & 2.25 & 0.910 (0.286) & 13.60 (6.84) \\
\bottomrule\end{tabular}}
\end{table}

\begin{table}[ht]\centering\scriptsize
\caption{Simulation of Model-1 with H = 20.}
\label{Table:M1H20}
\resizebox{\textwidth}{!}{%
\begin{tabular}{llccccc}\toprule
Method & $\eta_{T+h}^*$ & RMSE($\times 10^{-3}$) & MAPE(\%) & MDAE($\times 10^{-3}$) & Coverage & Length($\times 10^{-3}$) \\  \midrule
\multicolumn{7}{c}{$T = 250$} \\ [8pt]
M-H-GARCH ($p=5$) & $N(0,1)$ & 5.75 (4.08) & 53.45 (53.08) & 3.30 & 0.798 (0.402) & 14.80 (6.77) \\
 & $\hat F_\eta$ &  & &  & 0.822 (0.383) & 15.97 (7.85) \\
+ conformal & $N(0,1)$  &  & &  & 0.842 (0.365) & 16.03 (7.32) \\
+ conformal & $\hat F_\eta$  &  & &  & 0.856 (0.351) & 17.05 (8.31) \\[2pt]
ARMA-H-GARCH & $N(0,1)$ & \textbf{3.97} (2.85) & \textbf{33.31} (28.83) & \textbf{2.11} & 0.802 (0.399) & \textbf{8.79} (2.40) \\
 & $\hat F_\eta$ &  & &  & 0.860 (0.347) & 10.00 (3.18) \\
+ conformal  & $N(0,1)$ &  & &  & 0.864 (0.343) & 10.05 (4.02) \\
+ conformal & $\hat F_\eta$  &  & &  & \textbf{0.882} (0.323) & 10.77 (4.08) \\[3pt]
GARCH(1,1) & $N(0,1)$ & 4.34 (3.18) & 37.06 (34.12) & 2.31 & 0.880 (0.325) & 13.89 (8.26) \\[3pt]
EGARCH(1,1) & $N(0,1)$ & 4.05 (2.90) & 35.58 (32.65) & 2.21 & 0.858 (0.349) & 11.00 (3.97) \\[3pt]
naive LSTM ($p=5$) & $N(0,1)$ & 5.18 (3.69) & 47.33 (47.48) & 2.89 & 0.784 (0.412) & 14.51 (9.77) \\[5pt]
\multicolumn{7}{c}{$T = 500$} \\ [8pt]
M-H-GARCH ($p=5$) & $N(0,1)$ & 4.82 (2.91) & 52.62 (49.86) & 3.40 & 0.854 (0.353) & 16.76 (6.34) \\
 & $\hat F_\eta$ &  & &  & 0.872 (0.334) & 17.65 (6.29) \\
+ conformal & $N(0,1)$  &  & &  & 0.894 (0.308) & 18.01 (6.58) \\
+ conformal & $\hat F_\eta$  &  & &  & 0.906 (0.292) & 18.64 (6.48) \\[3pt]
ARMA-H-GARCH & $N(0,1)$ & \textbf{3.29} (2.25) & \textbf{29.89} (29.03) & \textbf{1.87} & 0.864 (0.343) & \textbf{8.93} (1.81) \\
 & $\hat F_\eta$ &  & &  & 0.910 (0.286) & 10.42 (2.23) \\
+ conformal & $N(0,1)$  &  & &  & 0.910 (0.286) & 10.69 (3.81) \\
+ conformal & $\hat F_\eta$  &  & &  & \textbf{0.928} (0.259) & 11.53 (3.62) \\[3pt]
GARCH(1,1) & $N(0,1)$ & 3.39 (2.20) & 33.65 (32.64) & 2.24 & 0.906 (0.292) & 12.90 (4.00) \\[3pt]
EGARCH(1,1) & $N(0,1)$ & 3.33 (2.18) & 32.90 (32.05) & 2.10 & 0.890 (0.313) & 10.56 (1.98) \\[3pt]
naive LSTM ($p=5$) & $N(0,1)$ & 4.73 (3.04) & 48.94 (48.12) & 2.90 & 0.844 (0.363) & 17.34 (9.79) \\
\bottomrule\end{tabular}}\end{table}

\begin{table}[ht]\centering\scriptsize
\caption{Simulation of Model-2 with H = 20.}
\label{Table:M2H20}
\resizebox{\textwidth}{!}{%
\begin{tabular}{llccccc}\toprule
Method & $\eta_{T+h}^*$ & RMSE($\times 10^{-3}$) & MAPE(\%) & MDAE($\times 10^{-3}$) & Coverage & Length($\times 10^{-3}$) \\ \midrule
\multicolumn{7}{c}{$T = 250$} \\ [8pt]
M-H-GARCH ($p=5$) & $N(0,1)$ & 4.42 (3.07) & 38.09 (38.80) & 2.39 & 0.882 (0.323) & 12.25 (5.04) \\
 & $\hat F_\eta$ &  & &  & 0.898 (0.303) & 13.12 (5.47) \\
+ conformal & $N(0,1)$  &  & &  & 0.904 (0.295) & 13.00 (5.57) \\
+ conformal & $\hat F_\eta$  &  & &  & \textbf{0.912} (0.284) & 13.72 (5.89) \\[2pt]
ARMA-H-GARCH & $N(0,1)$ & 3.67 (2.67) & 25.94 (22.73) & \textbf{1.70} & 0.792 (0.406) & \textbf{7.30} (1.54) \\
 & $\hat F_\eta$ &  & &  & 0.844 (0.363) & 8.37 (2.12) \\
+ conformal & $N(0,1)$  &  & &  & 0.824 (0.381) & 8.43 (2.93) \\
+ conformal & $\hat F_\eta$  &  & &  & 0.858 (0.349) & 9.11 (2.84) \\[3pt]
GARCH(1,1) & $N(0,1)$ & 3.62 (2.61) & 26.39 (23.23) & 1.78 & 0.894 (0.308) & 10.99 (4.86) \\[3pt]
EGARCH(1,1) & $N(0,1)$ & \textbf{3.59} (2.61) & \textbf{25.74} (22.57) & \textbf{1.70} & 0.870 (0.337) & 8.80 (2.39) \\[3pt]
naive LSTM ($p=5$) & $N(0,1)$ & 3.90 (2.81) & 30.32 (30.30) & 1.84 & 0.850 (0.357) & 10.78 (7.12) \\[5pt]
\multicolumn{7}{c}{$T = 500$} \\ [8pt]
M-H-GARCH ($p=5$) & $N(0,1)$ & 5.31 (4.03) & 39.72 (39.34) & 2.63 & 0.902 (0.298) & 14.17 (5.41) \\
 & $\hat F_\eta$ &  & &  & 0.922 (0.268) & 15.27 (5.88) \\
+ conformal & $N(0,1)$  &  & &  & 0.928 (0.259) & 15.09 (6.09) \\
 + conformal & $\hat F_\eta$ &  & &  & \textbf{0.942} (0.234) & 15.88 (6.17) \\[3pt]
ARMA-H-GARCH & $N(0,1)$ & 4.87 (3.98) & 26.68 (19.97) & 1.94 & 0.804 (0.397) & \textbf{7.74} (1.61) \\
 & $\hat F_\eta$ &  & &  & 0.864 (0.343) & 9.22 (2.58) \\
+ conformal & $N(0,1)$  &  & &  & 0.858 (0.349) & 10.11 (4.79) \\
+ conformal & $\hat F_\eta$  &  & &  & 0.880 (0.325) & 10.68 (4.42) \\[3pt]
GARCH(1,1) & $N(0,1)$ & \textbf{4.75} (3.94) & 25.65 (19.97) & \textbf{1.79} & 0.922 (0.268) & 11.10 (3.92) \\[3pt]
EGARCH(1,1) & $N(0,1)$ & 4.81 (3.99) & \textbf{25.49} (19.60) & 1.83 & 0.868 (0.339) & 8.86 (1.86) \\[3pt]
naive LSTM ($p=5$) & $N(0,1)$ & 5.29 (4.15) & 36.04 (37.64) & 2.30 & 0.892 (0.311) & 14.40 (9.30) \\
\bottomrule\end{tabular}}\end{table}

\FloatBarrier
\section{Real-data analysis}\label{Sec:realdata}
\begin{figure}[htbp]
    \centering
    \includegraphics[width=0.8\linewidth]{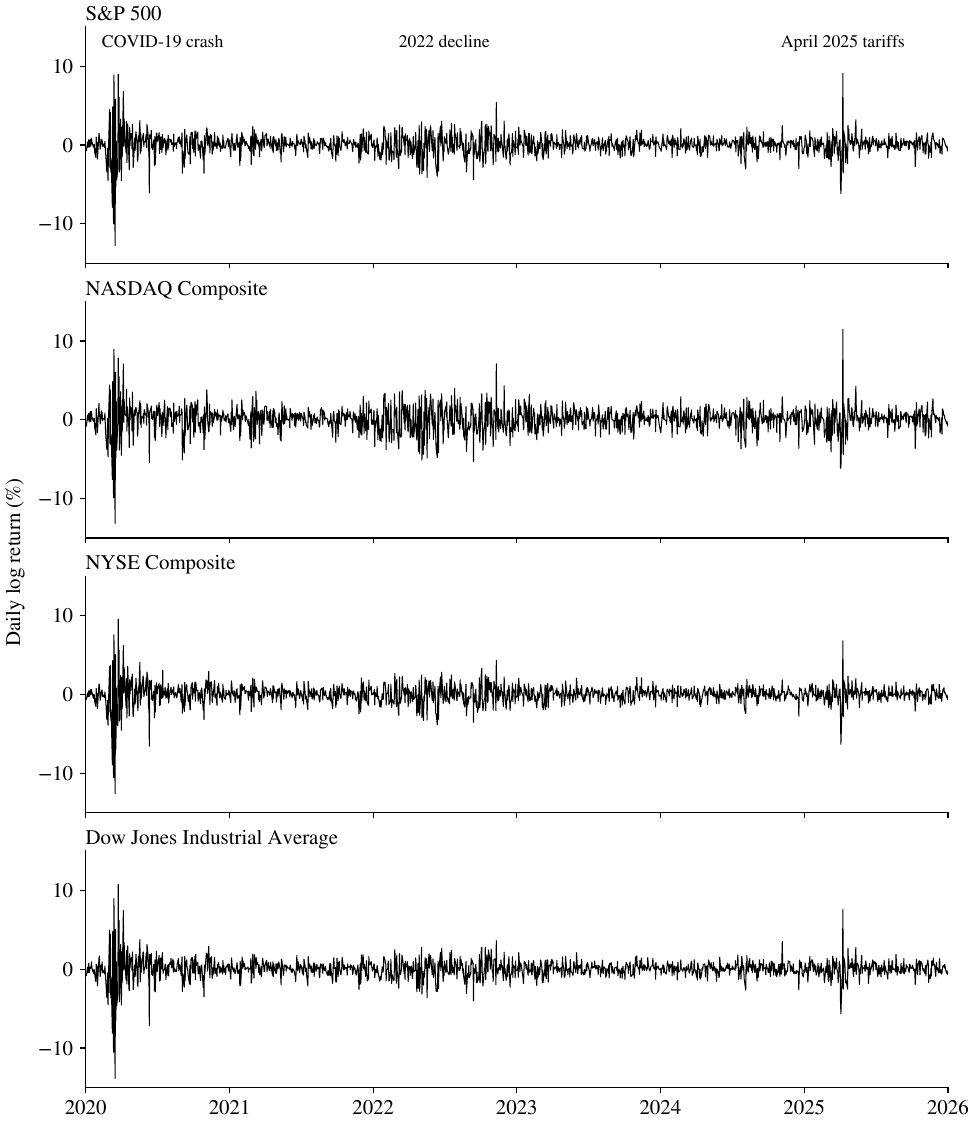}
    \caption{Log-returns of four real series from January 2, 2020 to December 31, 2025.}
    \label{fig:logreturn}
\end{figure}

Beyond the simulations, we consider multiple real datasets to compare the different methods applied in the simulation section. The closing prices of four U.S. stock indices, i.e., S\&P 500, the NASDAQ Composite, the NYSE Composite, and the Dow Jones Industrial Average, from January 2, 2020 to December 31, 2025 are downloaded from Yahoo Finance. In total, we have 1508 closing prices providing 1507 log-returns, which are defined as $y_t = \log(P_t / P_{t-1})$,\footnote{$P_t$ represents the closing price at the time index $t$} on the same trading dates for all four series. This time period was selected because it includes the COVID-19 crash of March 2020, when all four indices had their largest one-day fall (log returns of $-12.6\%$ to $-13.8\%$ on March 16, 2020), the 2022 market decline and the tariff-related swings of April 2025. The log-returns for four series are plotted in Figure \ref{fig:logreturn}. By looking at the graphs, it is clear that volatility clustering happens in early 2020, throughout 2022, and April 2025. These volatile series are deployed to challenge our methods on point predictions and constructing meaningful prediction intervals. 

To evaluate the performance of different methods, we take a rolling window prediction with window sizes of 250 (1-year) and 500 (2-year) to forecast $H = 5$ and $H = 20$ step-ahead realized volatility defined in Eq.~\eqref{Eq:futureVolatility}. More specifically, we do 1238
rolling window predictions per series for $T=250$ window size and 988 predictions for $T=500$ window size. Since we have two rolling window sizes (250/500), and two prediction horizons (5/20), we have 16 prediction scenarios in total for 4 real series. For each prediction scenario, we compute RMSE, MAPE and MDAE for all point predictions over the rolling-window predictions. Since RMSE, MAPE and MDAE differ in scale across scenarios, we report the relative performance of different methods to the benchmark GARCH(1,1). More specifically, for a metric with value $m_{k,s}$ from method $k$ in scenario $s$, the relative value of method $k$ is defined as the geometric mean $\exp\{\frac{1}{16}\sum_{s=1}^{16}\log(m_{k,s}/m_{\mathrm{GARCH},s})\}$; where $m_{\mathrm{GARCH},s}$ is the metric from the GARCH method in scenario $s$. Particularly, a relative value below $1$ means that the corresponding metric is smaller than that of GARCH(1,1) on average over the $16$ scenarios, and GARCH(1,1) itself has a value of $1$.  All relative values for RMSE, MAPE and MDAE are denoted by Rel-RMSE, Rel-MAPE and Rel-MDAE, respectively; see Table \ref{tab:real-overall-point} for details.

Besides the point prediction metrics, we also summarize the Mean-coverage as a proportion of the 95\% prediction interval that contains the future realized volatility over all 16 prediction scenarios. Also, we consider $|\mathrm{Cov} - 0.95|$, which is the average distance between an interval’s coverage and the 0.95 target over all 16 scenarios. In addition, we give the metric Min-coverage, which is the interval’s lowest empirical coverage across the 16 scenarios. Since the interval length is scale dependent, the relative values, namely Rel-length, of different methods are also provided against the benchmark GARCH(1,1) method; see Table \ref{tab:real-overall-pi} for details.

All in all, from the point prediction perspective, the ARMA-H-GARCH method dominates other variants. From the coverage of various prediction intervals, the prediction interval from the M-H-GARCH method with conformal extension and $\eta_{T+h}^*$ generated from $\hat F_\eta$ generally outperforms other variants. 

Integrating the simulation and real data results, we would recommend the ARMA-H-GARCH method for the point prediction tasks and the prediction interval based on M-H-GARCH with conformal extension and $\eta_{T+h}^*$ generated from $\hat F_\eta$ to build the prediction intervals.

\begin{table}[ht]\centering\small
\caption{Relative performance of different methods compared to GARCH(1,1) on point predictions}
\label{tab:real-overall-point}
\begin{tabular}{lccc}\toprule
Method & Rel-RMSE & Rel-MAPE & Rel-MDAE \\ \midrule
M-H-GARCH & 1.111 & 1.192 & 1.158 \\
ARMA-H-GARCH & \textbf{0.952} & \textbf{0.919} & \textbf{0.942} \\
GARCH(1,1) & 1.000 & 1.000 & 1.000 \\
EGARCH(1,1) & 0.956 & 0.923 & \textbf{0.942} \\
naive LSTM & 1.394 & 1.718 & 1.747 \\
\bottomrule\end{tabular}
\end{table}

\begin{table}[ht]\centering\small
\caption{Relative performance of different methods compared to GARCH(1,1) on performance of prediction intervals.}
\label{tab:real-overall-pi}
\begin{tabular}{llcccc}\toprule
Method & $\eta_{T+h}^*$ & Mean-coverage & $|\mathrm{Cov} - 0.95|$ & Min-coverage & Rel-length \\ \midrule
M-H-GARCH & $N(0,1)$ & 0.911 & 0.039 & 0.876 & 1.341 \\
 & $\hat F_\eta$ & 0.919 & 0.031 & 0.889 & 1.405 \\
 + conformal & $N(0,1)$  & 0.920 & 0.030 & 0.891 & 1.390 \\
 + conformal  & $\hat F_\eta$ & \textbf{0.925} & \textbf{0.025} & \textbf{0.898} & 1.437 \\[3pt]
ARMA-H-GARCH & $N(0,1)$ & 0.856 & 0.094 & 0.754 & \textbf{0.837} \\
 & $\hat F_\eta$ & 0.877 & 0.073 & 0.779 & 0.907 \\
+ conformal & $N(0,1)$  & 0.874 & 0.076 & 0.771 & 0.934 \\
+ conformal & $\hat F_\eta$  & 0.886 & 0.064 & 0.787 & 0.972 \\[3pt]
GARCH(1,1) & $N(0,1)$ & 0.885 & 0.065 & 0.793 & 1.000 \\[3pt]
EGARCH(1,1) & $N(0,1)$ & 0.853 & 0.097 & 0.767 & 0.864 \\[3pt]
naive LSTM & $N(0,1)$ & 0.856 & 0.094 & 0.782 & 1.784 \\
\bottomrule\end{tabular}
\end{table}

\FloatBarrier
\section{Conclusions}
\label{conclusions}
In this work, we give concrete algorithms based on forward bootstrap construction to make point predictions and build prediction intervals for future realized volatility. In summary, the hybrid LSTM model with the additional information from the estimated variance and coefficients of classical models has a better point prediction performance compared to naive LSTM and classical models. Moreover, improved by our two strategies, the resulting prediction interval has a good coverage on simulation and real data. 

However, it is clear that our prediction interval still undercovers future volatility even after being widened by conformal prediction. The reason for this is partially illustrated in Remark \ref{Remark:effectsofmodelestimation}, i.e., the estimation variability of our model is still not fully captured. A more conservative way to further relieve this undercoverage issue is making the so-called pertinent prediction interval by approximating the predictive root distribution with bootstrap. However, this procedure is computationally heavy due to the training procedure needing to be performed on each bootstrap data. Recently, \cite{politis2026pertinent} proposed a hybrid resampling scheme to construct pertinent prediction intervals in a computationally efficient way. A future extension of this work could leverage this hybrid resampling scheme to further improve the coverage of prediction intervals from the hybrid model. Also, the increasing availability of high-frequency financial data has spurred the development of new methods in financial economics. This high-frequency data analysis has wide application in financial economics; see \cite{fan2012vast} for portfolio selection, \cite{fan2008spot} for instantaneous volatility estimation, and \cite{russell2010analysis} for an analysis of the mechanisms through which prices reach equilibrium. It would be interesting to see how the hybrid models proposed in this work handle high-frequency data. Moreover, it is also intriguing to see how the hybrid model can take advantage of the high-frequency volatility models, e.g., parallel GARCH and realized GARCH models developed by \cite{shephard2010realising,hansen2012realized}.

\section{Acknowledgements}
There is no conflict of interest between the two authors. This work is partially supported by the Provost Fellowship of Loyola University Chicago.

\FloatBarrier

\clearpage
\bibliographystyle{apalike}
\bibliography{refs}

@article{schmidhuber1997lstm,
  title     = {Long Short-Term Memory},
  author    = {Hochreiter, Sepp and Schmidhuber, J{\"u}rgen},
  journal   = {Neural Computation},
  volume    = {9},
  number    = {8},
  pages     = {1735--1780},
  year      = {1997},
  publisher = {MIT Press},
  doi       = {10.1162/neco.1997.9.8.1735}
}

@article{Zhang2020,
  title     = {Physics-Informed Multi-{LSTM} Networks for Metamodeling of Nonlinear Structures},
  author    = {Zhang, Ruiyang and Liu, Yang and Sun, Hao},
  journal   = {Computer Methods in Applied Mechanics and Engineering},
  volume    = {369},
  pages     = {113226},
  year      = {2020},
  publisher = {Elsevier},
  doi       = {10.1016/j.cma.2020.113226}
}

@article{mandis2024,
  title     = {Exploring Nontraditional {LSTM} Architectures for Modeling Demethanizer Column Operations},
  author    = {Mandis, Marta and Baratti, Roberto and Chebeir, Jorge and Tronci, Stefania and Romagnoli, Jos{\'e} A.},
  journal   = {Computers \& Chemical Engineering},
  volume    = {183},
  pages     = {108591},
  year      = {2024},
  publisher = {Elsevier}
}

@article{Nelson1991EGARCH,
  title     = {Conditional Heteroskedasticity in Asset Returns: A New Approach},
  author    = {Nelson, Daniel B.},
  journal   = {Econometrica},
  volume    = {59},
  number    = {2},
  pages     = {347--370},
  year      = {1991}
}

@article{Roberts1959,
  title     = {Control Chart Tests Based on Geometric Moving Averages},
  author    = {Roberts, S. W.},
  journal   = {Technometrics},
  volume    = {1},
  number    = {3},
  pages     = {239--250},
  year      = {1959}
}

@article{Moghar2020,
  title     = {Stock Market Prediction Using {LSTM} Recurrent Neural Network},
  author    = {Moghar, Adil and Hamiche, Mhamed},
  journal   = {Procedia Computer Science},
  volume    = {170},
  pages     = {1168--1173},
  year      = {2020},
  publisher = {Elsevier}
}

@article{kim2018forecasting,
  title     = {Forecasting the Volatility of Stock Price Index: A Hybrid Model Integrating {LSTM} with Multiple {GARCH}-Type Models},
  author    = {Kim, Ha Young and Won, Chang Hyun},
  journal   = {Expert Systems with Applications},
  volume    = {103},
  pages     = {25--37},
  year      = {2018},
  publisher = {Elsevier}
}

@article{bollerslev1986garch,
  title     = {Generalized Autoregressive Conditional Heteroskedasticity},
  author    = {Bollerslev, Tim},
  journal   = {Journal of Econometrics},
  volume    = {31},
  number    = {3},
  pages     = {307--327},
  year      = {1986}
}

@article{garciamedina2023,
  title     = {{LSTM}--{GARCH} Hybrid Model for the Prediction of Volatility in Cryptocurrency Portfolios},
  author    = {Garc{\'\i}a-Medina, Andr{\'e}s and Aguayo-Moreno, Ester},
  journal   = {Computational Economics},
  volume    = {63},
  number    = {4},
  pages     = {1511--1542},
  year      = {2024},
  publisher = {Springer},
  doi       = {10.1007/s10614-023-10373-8}
}

@article{wanglu2024,
  title     = {{COMEX} Copper Futures Volatility Forecasting: Econometric Models and Deep Learning},
  author    = {Wang, Zian and Lu, Xinyi},
  journal   = {arXiv preprint arXiv:2409.08356},
  year      = {2024}
}

@article{glosten1993,
  title     = {On the Relation between the Expected Value and the Volatility of the Nominal Excess Return on Stocks},
  author    = {Glosten, Lawrence R. and Jagannathan, Ravi and Runkle, David E.},
  journal   = {The Journal of Finance},
  volume    = {48},
  number    = {5},
  pages     = {1779--1801},
  year      = {1993}
}

@techreport{longerstaey1996riskmetricstm,
  title       = {{RiskMetrics}---Technical Document},
  author      = {Longerstaey, Jacques and Spencer, Martin},
  institution = {Morgan Guaranty Trust Company of New York},
  address     = {New York},
  note        = {4th edition},
  year        = {1996}
}

@article{engle1982arch,
  title     = {Autoregressive Conditional Heteroscedasticity with Estimates of the Variance of {United Kingdom} Inflation},
  author    = {Engle, Robert F.},
  journal   = {Econometrica},
  volume    = {50},
  number    = {4},
  pages     = {987--1007},
  year      = {1982}
}

@article{mcculloch1943logical,
  title     = {A Logical Calculus of the Ideas Immanent in Nervous Activity},
  author    = {McCulloch, Warren S. and Pitts, Walter},
  journal   = {The Bulletin of Mathematical Biophysics},
  volume    = {5},
  number    = {4},
  pages     = {115--133},
  year      = {1943},
  publisher = {Springer}
}

@article{cybenko1989approximation,
  title     = {Approximation by Superpositions of a Sigmoidal Function},
  author    = {Cybenko, George},
  journal   = {Mathematics of Control, Signals and Systems},
  volume    = {2},
  number    = {4},
  pages     = {303--314},
  year      = {1989},
  publisher = {Springer}
}

@article{hornik1989multilayer,
  title     = {Multilayer Feedforward Networks Are Universal Approximators},
  author    = {Hornik, Kurt and Stinchcombe, Maxwell and White, Halbert},
  journal   = {Neural Networks},
  volume    = {2},
  number    = {5},
  pages     = {359--366},
  year      = {1989},
  publisher = {Elsevier}
}

@inproceedings{yarotsky2018optimal,
  title     = {Optimal Approximation of Continuous Functions by Very Deep {ReLU} Networks},
  author    = {Yarotsky, Dmitry},
  booktitle = {Proceedings of the 31st Conference on Learning Theory},
  series    = {Proceedings of Machine Learning Research},
  volume    = {75},
  pages     = {639--649},
  year      = {2018},
  publisher = {PMLR}
}

@article{yarotsky2020phase,
  title     = {The Phase Diagram of Approximation Rates for Deep Neural Networks},
  author    = {Yarotsky, Dmitry and Zhevnerchuk, Anton},
  journal   = {Advances in Neural Information Processing Systems},
  volume    = {33},
  pages     = {13005--13015},
  year      = {2020}
}

@article{hajizadeh2012hybrid,
  title     = {A Hybrid Modeling Approach for Forecasting the Volatility of {S\&P} 500 Index Return},
  author    = {Hajizadeh, Ehsan and Seifi, Abbas and {Fazel Zarandi}, M. H. and Turksen, I. B.},
  journal   = {Expert Systems with Applications},
  volume    = {39},
  number    = {1},
  pages     = {431--436},
  year      = {2012},
  publisher = {Elsevier}
}

@article{hansen2005forecast,
  title     = {A Forecast Comparison of Volatility Models: Does Anything Beat a {GARCH(1,1)}?},
  author    = {Hansen, Peter R. and Lunde, Asger},
  journal   = {Journal of Applied Econometrics},
  volume    = {20},
  number    = {7},
  pages     = {873--889},
  year      = {2005},
  publisher = {Wiley}
}

@book{francq2019garch,
  title     = {{GARCH} Models: Structure, Statistical Inference and Financial Applications},
  author    = {Francq, Christian and Zako{\"\i}an, Jean-Michel},
  edition   = {2},
  year      = {2019},
  publisher = {John Wiley \& Sons}
}

@article{politis2023multi,
  title     = {Multi-Step-Ahead Prediction Intervals for Nonparametric Autoregressions via Bootstrap: Consistency, Debiasing, and Pertinence},
  author    = {Politis, Dimitris N. and Wu, Kejin},
  journal   = {Stats},
  volume    = {6},
  number    = {3},
  pages     = {839--867},
  year      = {2023},
  publisher = {MDPI}
}

@article{wu2024bootstrap,
  title     = {Bootstrap Prediction Inference of Nonlinear Autoregressive Models},
  author    = {Wu, Kejin and Politis, Dimitris N.},
  journal   = {Journal of Time Series Analysis},
  volume    = {45},
  number    = {5},
  pages     = {800--822},
  year      = {2024},
  publisher = {Wiley}
}

@article{pan2016bootstrap,
  title     = {Bootstrap Prediction Intervals for Linear, Nonlinear and Nonparametric Autoregressions},
  author    = {Pan, Li and Politis, Dimitris N.},
  journal   = {Journal of Statistical Planning and Inference},
  volume    = {177},
  pages     = {1--27},
  year      = {2016},
  publisher = {Elsevier}
}

@book{politis2015model,
  title     = {Model-Free Prediction and Regression: A Transformation-Based Approach to Inference},
  author    = {Politis, Dimitris N.},
  series    = {Frontiers in Probability and the Statistical Sciences},
  publisher = {Springer},
  address   = {Cham},
  year      = {2015}
}

@article{wang2026model,
  title     = {Model-Free Bootstrap and Conformal Prediction in Regression: Conditionality, Conjecture Testing, and Pertinent Prediction Intervals},
  author    = {Wang, Yiren and Politis, Dimitris N.},
  journal   = {Journal of Nonparametric Statistics},
  volume    = {38},
  number    = {1},
  pages     = {311--345},
  year      = {2026},
  publisher = {Taylor \& Francis}
}

@article{wu2024deep,
  title     = {Deep Limit Model-Free Prediction in Regression},
  author    = {Wu, Kejin and Politis, Dimitris N.},
  journal   = {arXiv preprint arXiv:2408.09532},
  year      = {2024}
}

@article{wu2026prediction,
  title     = {Prediction Inference of Time Series with Standard {ReLU} Deep Neural Networks},
  author    = {Wu, Kejin},
  journal   = {arXiv preprint arXiv:2608.15362},
  year      = {2026}
}

@book{vovk2005algorithmic,
  title     = {Algorithmic Learning in a Random World},
  author    = {Vovk, Vladimir and Gammerman, Alexander and Shafer, Glenn},
  publisher = {Springer},
  address   = {New York},
  year      = {2005}
}

@article{sesia2021conformal,
  title     = {Conformal Prediction Using Conditional Histograms},
  author    = {Sesia, Matteo and Romano, Yaniv},
  journal   = {Advances in Neural Information Processing Systems},
  volume    = {34},
  pages     = {6304--6315},
  year      = {2021}
}

@article{romano2019conformalized,
  title     = {Conformalized Quantile Regression},
  author    = {Romano, Yaniv and Patterson, Evan and Cand{\`e}s, Emmanuel},
  journal   = {Advances in Neural Information Processing Systems},
  volume    = {32},
  year      = {2019}
}

@article{lei2018distribution,
  title     = {Distribution-Free Predictive Inference for Regression},
  author    = {Lei, Jing and G'Sell, Max and Rinaldo, Alessandro and Tibshirani, Ryan J. and Wasserman, Larry},
  journal   = {Journal of the American Statistical Association},
  volume    = {113},
  number    = {523},
  pages     = {1094--1111},
  year      = {2018},
  publisher = {Taylor \& Francis}
}

@article{chernozhukov2021distributional,
  title     = {Distributional Conformal Prediction},
  author    = {Chernozhukov, Victor and W{\"u}thrich, Kaspar and Zhu, Yinchu},
  journal   = {Proceedings of the National Academy of Sciences},
  volume    = {118},
  number    = {48},
  pages     = {e2107794118},
  year      = {2021},
  publisher = {National Academy of Sciences}
}

@article{wang2009nonlinear,
  title     = {Nonlinear Neural Network Forecasting Model for Stock Index Option Price: Hybrid {GJR}--{GARCH} Approach},
  author    = {Wang, Yi-Hsien},
  journal   = {Expert Systems with Applications},
  volume    = {36},
  number    = {1},
  pages     = {564--570},
  year      = {2009},
  publisher = {Elsevier}
}

@article{dai2026distributional,
  title     = {Distributional Conformal Prediction for {Markov} Processes},
  author    = {Dai, Dehao and Wu, Kejin and Politis, Dimitris N.},
  journal   = {arXiv preprint arXiv:2605.24848},
  year      = {2026}
}

@article{osg07,
  title     = {The {Open Science Grid}},
  author    = {Pordes, Ruth and Petravick, Don and Kramer, Bill and Olson, Doug and Livny, Miron and Roy, Alain and Avery, Paul and Blackburn, Kent and Wenaus, Torre and W{\"u}rthwein, Frank and Foster, Ian and Gardner, Rob and Wilde, Mike and Blatecky, Alan and McGee, John and Quick, Rob},
  journal   = {Journal of Physics: Conference Series},
  volume    = {78},
  number    = {1},
  pages     = {012057},
  year      = {2007},
  doi       = {10.1088/1742-6596/78/1/012057}
}

@inproceedings{osg09,
  title     = {The Pilot Way to Grid Resources Using {glideinWMS}},
  author    = {Sfiligoi, Igor and Bradley, Daniel C. and Holzman, Burt and Mhashilkar, Parag and Padhi, Sanjay and W{\"u}rthwein, Frank},
  booktitle = {2009 WRI World Congress on Computer Science and Information Engineering},
  volume    = {2},
  pages     = {428--432},
  year      = {2009},
  publisher = {IEEE},
  doi       = {10.1109/CSIE.2009.950}
}

@misc{osg2,
  title     = {{OSPool}},
  author    = {{OSG}},
  publisher = {OSG},
  year      = {2006},
  doi       = {10.21231/906P-4D78},
  url       = {https://osg-htc.org/services/open_science_pool.html}
}

@misc{osg1,
  title     = {{Open Science Data Federation}},
  author    = {{OSG}},
  publisher = {OSG},
  year      = {2015},
  doi       = {10.21231/0KVZ-VE57},
  url       = {https://osdf.osg-htc.org/}
}

@article{politis2026pertinent,
  title     = {Pertinent Prediction Intervals in Linear Regression},
  author    = {Politis, Dimitris N.},
  journal   = {Stats},
  volume    = {9},
  number    = {4},
  pages     = {68},
  year      = {2026},
  publisher = {MDPI}
}

@article{fan2012vast,
  title     = {Vast Volatility Matrix Estimation Using High-Frequency Data for Portfolio Selection},
  author    = {Fan, Jianqing and Li, Yingying and Yu, Ke},
  journal   = {Journal of the American Statistical Association},
  volume    = {107},
  number    = {497},
  pages     = {412--428},
  year      = {2012},
  publisher = {Taylor \& Francis}
}

@article{fan2008spot,
  title     = {Spot Volatility Estimation for High-Frequency Data},
  author    = {Fan, Jianqing and Wang, Yazhen},
  journal   = {Statistics and Its Interface},
  volume    = {1},
  number    = {2},
  pages     = {279--288},
  year      = {2008},
  publisher = {International Press of Boston}
}

@incollection{russell2010analysis,
  title     = {Analysis of High-Frequency Data},
  author    = {Russell, Jeffrey R. and Engle, Robert F.},
  booktitle = {Handbook of Financial Econometrics: Tools and Techniques},
  editor    = {A{\"\i}t-Sahalia, Yacine and Hansen, Lars Peter},
  pages     = {383--426},
  year      = {2010},
  publisher = {Elsevier}
}

@article{shephard2010realising,
  title     = {Realising the Future: Forecasting with High-Frequency-Based Volatility ({HEAVY}) Models},
  author    = {Shephard, Neil and Sheppard, Kevin},
  journal   = {Journal of Applied Econometrics},
  volume    = {25},
  number    = {2},
  pages     = {197--231},
  year      = {2010},
  publisher = {Wiley}
}

@article{hansen2012realized,
  title     = {Realized {GARCH}: A Joint Model for Returns and Realized Measures of Volatility},
  author    = {Hansen, Peter R. and Huang, Zhuo and Shek, Howard Howan},
  journal   = {Journal of Applied Econometrics},
  volume    = {27},
  number    = {6},
  pages     = {877--906},
  year      = {2012},
  publisher = {Wiley}
}

\appendix

\section{Implementation details}\label{sec:implementation}
In this section, we provide more implementation details and the training policies for various methods. Every method involving a network (M-H-GARCH, ARMA-H-GARCH, and the naive LSTM) uses the same architecture: a single LSTM layer with $32$ hidden units, a sequence length of $5$, i.e., the 5 most recent inputs, and the Adam optimizer with a learning rate of $10^{-3}$. The naive LSTM thus differs from the hybrids only in its inputs, which consist of the return series alone without the parameter block $\Theta$ or the fitted GARCH, EGARCH and EWMA variance series. For these variance series, to improve numerical stability, they are truncated to $[10^{-12}, 1]$ whenever they are updated along a bootstrap pseudo series $\{y^*_{T+h}\}$. The single LSTM layer structure has been taken from previous studies \cite{kim2018forecasting,hajizadeh2012hybrid,wang2009nonlinear}, etc.

All networks are trained by full-batch Adam, and the number of training epochs is chosen by early stopping rather than fixed in advance. For the stopping rule, the last 20\% of the training sample, $t \geq \varpi$ (the same window used by Algorithm \ref{Algo:conformalextension}), serves as a validation window. More specifically, the network is first fitted on $t < \varpi$ while its training loss, (\ref{Eq:MSEloss}) for M-H-GARCH and the naive LSTM and (\ref{Eq:LLF}) for ARMA-H-GARCH, is evaluated on the validation window after every epoch. Training stops once the validation loss has not improved for 100 consecutive epochs, or at a maximum of 1200 epochs, and the number of epochs $e^*$ with the smallest validation loss is retained. The network is then refitted on the full training sample for $e^*$ epochs, starting from the same initialization.  The time-varying network inputs are standardized by their training-sample means and standard deviations, and the network output is $\hat\sigma_t = s\cdot\mathrm{softplus}(\cdot)$ for a fixed training-sample volatility level $s$, so that the returned $\hat\sigma_t$ is positive. 

We should also mention that ARMA-H-GARCH model is estimated in two stages. The ARMA coefficients $\alpha$ are first initialized by conditional maximum likelihood with a constant variance over 300 epochs; no network is trained at this stage, and the LSTM is constructed only afterward. The network weights and $\Pi$ are then updated together by minimizing loss (\ref{Eq:LLF}), with the number of joint epochs chosen by the same early-stopping rule (at most 900). Both stages are first run on $t < \varpi$ to select $e^*$, and then rerun on the full sample. Moreover, the partial autocorrelations of the AR and MA polynomials are constrained to $(-0.98, 0.98)$ by a $\tanh$ reparameterization and mapped to the coefficients by the Durbin--Levinson recursion, which keeps the recursion inside the stationarity and invertibility region for any order. 

The simulation studies are performed using computational services provided by the OSG Consortium \cite{osg07,osg09,osg2,osg1}.
To make our simulations reproducible, we used a fixed master seed ${\tt BASE\_SEED = 20261023 }$ from NumPy seed sequence. In particular, the seed for each replication with different simulation models is defined as ${\tt np.random.SeedSequence([20261023, model, T, rep])}$, where ${\tt model}$ stands for the simulation model; ${\tt T}$ stands for the sample size;  ${\tt rep}$ stands for the replication index.

\end{document}